\documentclass[11pt]{article}

\usepackage{amsmath,amssymb}
\usepackage{graphicx}
\usepackage{color}

\allowdisplaybreaks 

\newcommand{\dd}[2]{\frac{\mathrm{d} #1}{\mathrm{d} #2}}

\newcommand{\pds}[2]{\partial #1 / \partial #2}

\newcommand{\ddd}{\mathrm{d}}
\newcommand{\p}{\partial}

\newcommand{\todo}[1][\null]{\ensuremath{\clubsuit}}
\newcommand{\const}{\mathop{\rm const}\nolimits}

\newcommand{\lsemioplus}{\mathbin{\mbox{$\lefteqn{\hspace{.77ex}\rule{.4pt}{1.2ex}}{\in}$}}}

\newcommand{\vv}{\mathbf{v}}
\newcommand{\ww}{\mathbf{w}}

\newcommand{\nn}{\mathbf{\nabla}}
\newcommand{\ve}{\varepsilon}
\newcommand{\Ad}{\mathrm{Ad}}

\newcommand{\ZZ}{\mathcal{Z}}
\newcommand{\JJ}{\mathcal{J}}
\newcommand{\XX}{\mathcal{X}}
\newcommand{\YY}{\mathcal{Y}}
\newcommand{\DDD}{\mathcal{D}}

\newcommand{\bbve}{$\beta$BVE}
\newcommand{\sbve}{sBVE}

\begin{document}

\par\noindent {\LARGE\bf
Lie symmetries and exact solutions \\ of the barotropic vorticity equation
\par}
{\vspace{4mm}\par\noindent {\bf Alexander Bihlo~$^\dag$ and Roman O. Popovych~$^\dag\, ^\ddag$
} \par\vspace{2mm}\par}

{\vspace{2mm}\par\noindent {\it
$^{\dag}$~Faculty of Mathematics, University of Vienna, Nordbergstra{\ss}e 15, A-1090 Vienna, Austria\\
}}
{\noindent \vspace{2mm}{\it
$\phantom{^\dag}$~\textup{E-mail}: alexander.bihlo@univie.ac.at
}\par}

{\vspace{2mm}\par\noindent {\it
$^\ddag$~Institute of Mathematics of NAS of Ukraine, 3 Tereshchenkivska Str., 01601 Kyiv, Ukraine\\
}}
{\noindent \vspace{2mm}{\it
$\phantom{^\dag}$~\textup{E-mail}: rop@imath.kiev.ua
}\par}

\vspace{6mm}\par\noindent\hspace*{8mm}\parbox{140mm}{\small\looseness=-1 
Lie group methods are used for the study of various issues related to symmetries and exact solutions of the barotropic vorticity equation. 
The Lie symmetries of the barotropic vorticity equations 
on the $f$- and $\beta$-planes, as well as on the sphere 
in rotating and rest reference frames, are determined. 
A symmetry background for reducing the rotating reference frame to the rest frame is presented. 
The one- and two-dimensional inequivalent subalgebras of the Lie invariance algebras of both equations are exhaustively classified 
and then used to compute invariant solutions of the vorticity equations. 
This provides large classes of exact solutions, which include both Rossby and Rossby--Haurwitz waves as special cases. 
We also discuss the possibility of partial invariance for the $\beta$-plane equation, 
thereby further extending the family of its exact solutions. 
This is done in a more systematic and complete way than previously available in literature. 
}\par\vspace{6mm}

\section{Introduction}

The governing equations of geophysical fluid dynamics are mainly nonlinear partial differential equations (PDE). Since there is no general theory available for solving such equations, it is known to be very difficult to systematically construct their exact solutions. In meteorology, this problem is usually overcome by solving the governing equations numerically. However, as models become more sophisticated, it may be difficult to directly evaluate the quality of these numerical results. Moreover, it is dissatisfactory to rely solely on numerical modeling when studying the physics of the atmosphere. It is thus to be expected that exact solutions can both enhance our understanding of atmospheric processes and provide consistency tests for numerical models. 

The classical method of reduction of PDEs by using its Lie symmetries~\cite{olve86Ay,ovsi82Ay} and the extension to partially invariant solutions~\cite{ovsi82Ay} provides a manageable way to systematically construct exact solutions. It is the goal of this paper to carry out a comprehensive symmetry investigation of the barotropic vorticity equation both on the $\beta$-plane and on the sphere. Although there are already a number of works on the $\beta$-plane equation \cite{bihl07Ay,blen91Ay,huan04By,ibra95Ay,katk65Ay,katk66Ay}, none of them gives a systematic and complete symmetry analysis. In Ref.\ \cite{bihl07Ay} the classification of inequivalent subalgebras is done only for the one-dimensional case. The symmetry properties were used in Ref.\ \cite{blen91Ay} in order to obtain new solutions from the known ones. In a recent paper \cite{huan04By}, the procedure of group-invariant reduction is done without reference to the algebraic aspects of the classification problem. Consequently, some of the reductions presented in Ref.~\cite{huan04By} are overly complicated, and hence in some cases, these authors were only able to obtain some particular solutions (most notably of the well-known Rossby wave class). This reveals that the vorticity equation has classes of completely integrable reduced PDEs, as shown in the present paper. Finally, Refs.~\cite{katk65Ay,katk66Ay} (see also Ref.~\cite[pp.\ 221--225]{ibra95Ay}) also contain only a nonsystematic list of some group-invariant solutions. To the best of our knowledge, the spherical equation has not been investigated in light of its symmetries at all so far.

We divide this paper into two main parts: the first dealing with the symmetry analysis of the equation on the $\beta$-plane and the second considering the spherical version. For both equations, we determine the maximal Lie invariance algebras and classify their one- and two-dimensional subalgebras. Based on this classification, we give a complete list of group-invariant reduced equations and then demonstrate that Rossby (Rossby--Haurwitz) waves can be realized as group-invariant solutions of the barotropic vorticity equation on the plane (the sphere). Also, by means of algebraic inspection of the Lie symmetry algebras, it is shown that for the spherical equation there is no need to consider rotation of the Earth. Finally, some examples for partially invariant solutions will be given for the $\beta$-plane equation.

\section{The $\beta$-plane equation}

This section contains the classical symmetry analysis of the barotropic vorticity equation on the $\beta$-plane (\bbve).

\subsection{The model}

Assuming the two-dimensional velocity field $\vv$ to be nondivergent, 
it is possible to cast the Euler equations of an ideal fluid in a rotating reference frame as the conservation law of absolute vorticity $\eta = \zeta + f$, 
where $\zeta = \mathbf{k}\cdot(\nn\times\vv)$ is the vertical component of the vorticity vector (relative vorticity) 
and $f$ denotes the vertical Coriolis parameter, which depends only on~$y$. 
In what follows, we approximate $f$ by its truncated Taylor series, $f= f_0 + \ddd f / \ddd y|_{0}\, y =: f_0 + \beta y$, 
which leads to the $\beta$-plane approximation \cite{holt04Ay}. 
The Euler equations can then be equivalently written as the \bbve
\begin{align}\label{eq:vortbeta}
    \zeta_t + \psi_x\zeta_y-\psi_y\zeta_x+\beta\psi_x = 0,
\end{align}
where we have used the shorthand notation $\zeta_t=\pds{\zeta}{t}$, etc. 
The stream function $\psi=\psi(t,x,y)$ generates a nondivergent flow. 
It is related to the vorticity by means of the Laplacian, \[\zeta:=\psi_{xx}+\psi_{yy}.\] 
Rescaling allows us to set $\beta=1$, but for physical reasons this is not desired here.

\subsection{The symmetries}

The barotropic vorticity equation can be considered as a submodel of the ideal Euler equations, 
which have been thoroughly investigated in light of their symmetries (see, e.g., Refs.~\cite{andr98Ay,popo00Ay}).

Nevertheless, it is instructive to consider the symmetries of the barotropic vorticity equation separately to work out the peculiarities of large scale, two-dimensional fluid dynamics. 
It is quite common for different models of incompressible fluids that they admit infinite dimensional maximal Lie invariance algebras of a special structure. 
Techniques for handling such infinite-dimensional Lie algebras in order to solve hydrodynamic equations are given, e.g., in Ref.\ \cite{fush94Ay}. 
To the best of our knowledge, the symmetry algebra of the barotropic vorticity equation in the regular case $\beta\ne0$,
as well as some exact solutions, was first computed in Refs.\ \cite{katk65Ay,katk66Ay} (see also Refs.\ \cite{blen91Ay,ibra95Ay}). 
The fact that the singular case $\beta=0$ admits nontrivial symmetries has been known for a long time \cite{berk63Ay}. 
The corresponding maximal Lie symmetry algebra was rigorously calculated in Ref.~\cite{andr88Ay} 
(see also Ref.~\cite{andr98Ay}). 
It is significantly larger than for the regular case of~$\beta$.

We recomputed the symmetry algebras for our purposes and checked them with the computer algebra programs MuLie \cite{head93Ay} and DESOLV \cite{carm00Ay}. 
In the singular case $\beta=0$ corresponding to dynamics on the $f$-plane, the vorticity equation admits the infinite dimensional Lie symmetry algebra $\mathcal{B}^\infty_0$ with the basis generators
\begin{align*}
   &\DDD_1 = t\p_t - \psi\p_{\psi} &&\qquad \DDD_2 = 2\psi\p_{\psi} + x\p_x + y\p_y \\
   &\JJ = -y\p_x + x\p_y &&\qquad \JJ^t = -ty\p_x+tx\p_y + \tfrac{1}{2}(x^2+y^2)\p_{\psi} \\
   &\p_t &&\qquad \ZZ(g) = g(t)\p_{\psi} \\
   &\XX(f) = f(t)\p_x - f'(t)y\p_{\psi} &&\qquad \YY(h) = h(t)\p_y+h'(t)x\p_{\psi}
\end{align*}
where $h$, $f$ and $g$ run through the set of real-valued time-dependent functions. The shorthand notation for partial derivatives, e.g., $\p_t = \pds{}{t}$ is used. The physical significance of these generators is as follows: $\DDD_1$ and $\DDD_2$ are scaling operators, $\p_t$ generates time translations, and $\JJ$ and $\JJ^t$ correspond to rotations and time-dependent rotations in the horizontal plane. The operators $\mathcal{Y}(h)$ and $\mathcal{X}(f)$ are the infinitesimals of generalized transformations on a time-dependent moving coordinate system in the $y$- and $x$-directions, respectively. The generator $\mathcal{Z}(g)$ represents gauging the stream function. 

Likewise, in the case $\beta\ne 0$, eqn.~\eqref{eq:vortbeta} admits an infinite-dimensional Lie symmetry algebra $\mathcal{B}^\infty_1$, which is a subalgebra of $\mathcal{B}^\infty_0$. The basis generators of $\mathcal{B}^\infty_1$ are
\[
   \DDD = \DDD_1 - \DDD_2 = t\p_t - x\p_x - y\p_y - 3\psi\p_{\psi}, \qquad \p_t, \qquad \YY(1) = \p_y, \qquad \XX(f), \qquad \ZZ(g).
\]
The physical importance of this algebra is obvious from the $\beta=0$ case. As shown in the next section, this remarkable difference between the cases of vanishing and nonvanishing $\beta$ has no counterpart in spherical coordinates.

For the sake of completeness, we mention that the vorticity equation admits the discrete symmetries
$(t,x,y,\psi)\mapsto (-t,-x,y,\psi)$ and $(t,x,y,\psi)\mapsto (t,x,-y,-\psi)$ 
as well as their composition and their compositions with continuous symmetries.

\subsection{Classification of subalgebras}\label{sec:class2beta}

For an efficient and systematic computation of invariant solutions of PDEs, it is crucial to classify their Lie symmetry subalgebras. This is done upon using the adjoint action of a Lie group on its Lie algebra, which allows to determine the simplest representatives of equivalent subalgebras. The adjoint action of $\exp(\ve\vv)$ on $\ww_0$ is defined as the Lie series,
\[
  \ww(\ve) = \Ad(\exp(\ve\vv))\ww_0 := \sum_{n=0}^{\infty}\frac{\ve^n}{n!}\{\vv^n,\ww_0 \},
\]
where we introduced a shorthand notation for nested commutators: $\{\vv^0,\ww_0\} := \ww_0$, $\{\vv^n,\ww_0\} := (-1)^n[\vv,\{\vv^{n-1},\ww_0\}]$. Alternatively,~the adjoint representation can also be calculated by integrating the initial value problem 
\[
    \dd{\ww(\ve)}{\ve} = [\ww(\ve),\vv], \quad   \ww(0) = \ww_0.
\]
The nonidentical adjoint actions with basis elements of the algebra $\mathcal{B}^\infty_1$ are exhausted by the following list:
\begin{align*}
    &\Ad(e^{\ve\p_t})\DDD = \DDD - \ve\p_t && \Ad(e^{\ve\XX(v)})\DDD = \DDD+\ve\XX(v+tv')\\
    &\Ad(e^{\ve\p_y})\DDD = \DDD + \ve\p_y && \Ad(e^{\ve\ZZ(u)})\DDD = \DDD+\ve\ZZ(3u+tu')\\
    &\Ad(e^{\ve\DDD})\p_t = e^{\ve}\p_t &&\Ad(e^{\ve\ZZ(u)})\p_t = \p_t+\ve\ZZ(u') \\
    &\Ad(e^{\ve\XX(v)})\p_t = \p_t+\ve\XX(v') && \Ad(e^{\ve\DDD})\p_y = e^{-\ve}\p_y\\
    &\Ad(e^{\ve\XX(v)})\p_y = \p_y - \ve\ZZ(v') && \Ad(e^{\ve\DDD})\XX(v) = \XX(\tilde v), \quad \tilde v=e^{-\ve}v(e^{-\ve}t)\\
    &\Ad(e^{\ve\p_t})\XX(v) = \XX(\tilde v), \quad \tilde v=v(t-\ve) && \Ad(e^{\ve\p_y})\XX(f) = \XX(f)+\ZZ(f')\\
    &\Ad(e^{\ve\DDD})\ZZ(u) = \ZZ(\tilde u), \quad \tilde u=e^{-3\ve}u(e^{-\ve}t)&& \Ad(e^{\ve\p_t})\ZZ(u) = \ZZ(\tilde u), \quad \tilde u=u(t-\ve).
\end{align*}
They are used subsequently to classify the one- and two-dimensional subalgebras of~$\mathcal{B}^\infty_1$.

The approach to the classification of \textit{one-dimensional subalgebras} is fairly inductive: one takes the most general form of an infinitesimal generator of~$\mathcal{B}^\infty_1$,
$
    \vv = a_D\DDD + a_t\p_t + a_y\p_y + \XX(f) + \ZZ(g),
$
and subsequently tries to simplify it using adjoint actions and scaling by a nonvanishing constant multiplier \cite{olve86Ay}. This is done under additional assumptions on the constants $a_D$, $a_t$, and $a_y$ and the functions $f(t)$ and $g(t)$. Finally, the optimal set of conjugacy inequivalent one-dimensional subalgebras of~$\mathcal{B}^\infty_1$ reads
\begin{align}\label{eq:classificationbeta}
  \langle \DDD \rangle,\qquad \langle \p_t+c\p_y\rangle,\qquad \langle\p_y + \XX(f) \rangle,\qquad \langle \XX(f)+\ZZ(g) \rangle,
\end{align}
where $c\in\{0,\pm 1\}$. By means of using the discrete symmetry $(t,x,y,\psi)\mapsto(t,x,-y,-\psi)$ we can further assume $c\in\{0,1\}$. Moreover, due to adjoint actions, there are additional equivalences inside the third and fourth cases. In the third case, we can apply the adjoint actions $\Ad(e^{\ve\DDD})$ to rescale the argument $t$ and the function $f$ and $\Ad(e^{\ve\p_t})$ to shift the argument $t$ of the function $f$, respectively. 
In the fourth class, the additional equivalences are generated by $\Ad(e^{\ve\p_t})$,  $\Ad(e^{\ve\DDD})$, $\Ad(e^{\ve\p_y})$ and scaling the basis elements. So, the subalgebras $\langle \XX(f)+\ZZ(g) \rangle$ and  $\langle \XX(\tilde f)+\ZZ(\tilde g) \rangle$ are equivalent if and only if $\tilde f(t)=af(e^{\ve_2}t+\ve_1)$, $\tilde g(t)=ag(e^{\ve_2}t+\ve_1)+\ve_3f'(e^{\ve_2}t+\ve_1)$ for some constants $\ve_1$, $\ve_2$, $\ve_3$, and $a$, where $a\ne0$.


The procedure for the classification of \textit{two-dimensional subalgebras} is quite the same as in the one-dimensional case: one takes the most general form of two (linearly independent) infinitesimal generators
$
    \vv^i = a^i_D\DDD + a^i_t\p_t + a^i_y\p_y + \XX(f^i) + \ZZ(g^i)
$
with $i=1,2$ and tries to simultaneously cast them in a simpler form. This can be done by taking their nondegenerate linear combinations and acting on them by adjoint actions under different assumptions on the constants $\smash{a^i_D}$, $\smash{a^i_t}$, and $\smash{a^i_y}$ and/or functions $f^i$ and $g^i$. Since the generators $\vv^i$ form a subalgebra, one additionally has to ensure that the commutator of $\vv^1$ and $\vv^2$ lies in their span. This places further restrictions on both the constants $a^i_D$, $a^i_t$, and $a^i_y$ and on the functions $f^i(t)$, $g^i(t)$. 

Since the classification of two-dimensional subalgebras is somewhat lengthy, we only give the final result here. The list of inequivalent algebras reads
\begin{gather*}
    \left\langle \DDD,\ \p_t \right\rangle,\quad  \left\langle \DDD,\ \p_y + a\XX(1) \right\rangle,\quad  \left\langle \DDD,\ \XX(|t|^a) + c\ZZ(|t|^{a-2})\right\rangle,\quad  \left\langle \DDD,\ \ZZ(|t|^{a-2})\right\rangle, \\
    \left\langle \p_t + b\p_y,\ \XX(e^{a t})+ \ZZ((a bt+c)e^{a t})\right\rangle,\quad \left\langle \p_t + b\p_y,\ \ZZ((a bt+c)e^{a t})\right\rangle, \\
    \left\langle \p_y + \XX(f^1),\ \XX(1) + \ZZ(g^2)\right\rangle, \quad \left\langle \p_y + \XX(f^1),\ \ZZ(g^2)\right\rangle, \quad  \left\langle \XX(f^1)+\ZZ(g^1),\ \XX(f^2)+\ZZ(g^2)\right\rangle,
\end{gather*}
where $a,b$,and $c$ are arbitrary constants and $f^i=f^i(t)$ and $g^i=g^i(t)$, $i=1,2$, are arbitrary functions of time. In the last Lie subalgebra, the pairs of functions $(f^1,g^1)$ and $(f^2,g^2)$ have to be linearly independent. Again, within of the above classes there is an additional equivalence due to adjoint actions and changes of the basis.

Computing the associated group-invariant solutions based on the above classification 
determines those cases that have to be considered for receiving the complete set of inequivalent solutions.

\subsection{Group-invariant reduction with one-dimensional subalgebras} \label{sec:red1beta}

In what follows, we give the complete list of reduced equations obtained by imposing invariance under the one-parameter groups associated with the Lie algebras~\eqref{eq:classificationbeta}. In all four cases, $p,q$ denote the new independent variables, while $v=v(p,q)$ is the new dependent variable.
In the last case, an ansatz for $\psi$ exists only for the values of~$t$, where $f\ne 0$.

\begin{center}
\begin{tabular}{cp{4cm}p{4cm}c}
\hline
\rule[-3mm]{0mm}{8mm}{\bf 1} & $\langle \DDD \rangle$ & $\psi = t^{-3}v$ & $p = tx,\quad q = ty$ \\
\rule[-3mm]{0mm}{5mm}&\multicolumn{3}{c}{$-w +pw_p + qw_q + v_pw_q-v_qw_p + \beta v_p = 0, \quad w := v_{pp} + v_{qq}$}\\
\hline
\rule[-3mm]{0mm}{8mm}{\bf 2} & $\langle \p_t+c\p_y\rangle$ & $\psi = v$ & $p = x,\quad q = y-ct$ \\
\rule[-3mm]{0mm}{5mm}&\multicolumn{3}{c}{$-cw_q + v_pw_q - v_qw_p + \beta v_p = 0, \quad w := v_{pp} + v_{qq}$} \\
\hline
\rule[-3mm]{0mm}{8mm}{\bf 3} & $\langle\p_y + \XX(f) \rangle$ & $\psi = v - \frac{1}{2}f'y^2$ & $p = x-fy,\quad q = t$ \\
\rule[-3mm]{0mm}{5mm}&\multicolumn{3}{c}{$(1+f^2)w_{q} + 2ff'w + \beta v_p - f'' = 0, \quad w := v_{pp}$}\\ 
\hline
\rule[-3mm]{0mm}{8mm}{\bf 4} & $\langle \XX(f) +\ZZ(g)\rangle$ & $\psi = v - \frac{f'}{f}xy+\frac{g}{f}x$ & $p = y,\quad q = t$ \\
\rule[-3mm]{0mm}{5mm}&\multicolumn{3}{c}{$w_q + \left(\frac{g}{f}-\frac{f'}{f}p\right)w_p + \beta\left(\frac{g}{f}-\frac{f'}{f}p\right)=0, \quad 
w:=v_{pp}$}\\
\hline
\end{tabular}
\end{center}

We now discuss some properties and/or explicit solutions derived from the different cases considered above.

\medskip

\noindent Case~{\bf 1}. The reduced equation admits the maximal Lie invariance algebra $\langle \p_p+q\p_v,\, \p_v\rangle$. 
The basis operators $\p_p+q\p_v$ and $\p_v$ are induced by the operators $\XX(t^{-1})$ and $\ZZ(t^{-3})$ from $\mathcal{B}^\infty_1$, respectively.
Hence, there are no hidden symmetries related to this reduction.

\medskip

\noindent Case~{\bf 2}. The reduced equation admits the maximal Lie invariance algebra  
$\langle \p_p,\, \p_q,\, \p_v\rangle$ or $\langle p\p_p+q\p_q+3v\p_v,\, \p_p,\, \p_q,\, \p_v\rangle$ 
if $c=1$ or $c=0$, respectively. 
Any operator from this algebra is obviously induced by an operator from $\mathcal{B}^\infty_1$. 
For the basis operators, we have the following correspondence:
$\p_p\leftarrow\XX(1)$, $\p_q\leftarrow\p_y$, $\p_v\leftarrow\ZZ(1)$, $p\p_p+q\p_q+3v\p_v\leftarrow-\DDD$.
Again, we have no related hidden symmetries.

\medskip

As a result, further Lie reductions in both of the above cases give no new solutions in comparison to Lie reduction with respect to two-dimensional
subalgebras.

\medskip

\noindent Case {\bf 3} admits an exhaustive description for its general solution. Integrating once with respect to $p$ and using a change of coordinates yields
\begin{equation}\label{eq:kleingordon}
    \tilde v_{\tilde p\tilde q} + \beta \tilde v = 0,
\end{equation}
where
\[
\tilde v =  \frac v{1+f^2} - \frac{f''}{\beta}p + \frac h{\beta} + \frac{((1+f^2)f'')'}{\beta^2},\qquad \tilde q = \int \frac{\ddd q}{1+f^2},\qquad \tilde p = p,
\]
and $h=h(q)$ is an arbitrary smooth function of $q=t$.
Eqn.~\eqref{eq:kleingordon} is the one-dimensional Klein--Gordon equation (presented in the light-cone variables). 
For a list of exact solutions of this equation see, e.g.~\cite{poly02Ay}. 
It is straightforward to recover the famous Rossby wave solution upon using a harmonic ansatz for $\tilde v$ and choosing either $f=0$ 
(one-dimensional Rossby waves) or $f = \const$ (two-dimensional Rossby waves) \cite{bihl07Ay}.

\medskip

\noindent Case {\bf 4} is completely integrable by quadratures. First, we determine $w$ by solving the characteristic system. 
Afterwards, we integrate twice with respect to $p$ to determine $v$. 
Finally, substituting the expression so obtained for~$v$ into the ansatz for $\psi$, we arrive at the corresponding group-invariant solution
\[
    \psi = \frac{1}{f^2}F(\theta) - \frac{1}{6}\beta y^3 + h^1y + h^0 - \frac{f'}{f}xy+\frac{g}{f}x, 
\]
where $f$, $g$, $h^1$, and $h^0$ are arbitrary smooth functions of~$t$, $F$ is an arbitrary smooth function of $\theta=fy-\int g\,{\rm d}t$. 
The functions $h^1$ and $h^0$ can be set equal to 0 by symmetry transformations generated by an operator of the form $\XX(f)+\ZZ(g)$.

\subsection{Group-invariant reduction with two-dimensional subalgebras}\label{sec:red2beta}

Having considered reduction with one-dimensional subalgebras, it is not overly difficult to investigate reduction with two-dimensional subalgebras as well. Namely, the general solutions of cases {\bf 3} and {\bf 4} from section \ref{sec:red1beta} are completely described. That is, it is not necessary to consider reduction with two-dimensional subalgebras containing the generators $\p_y + \XX(f)$ and $\XX(f)+\ZZ(g)$. Moreover, since all algebras containing $\ZZ(g)$ cannot be used for a classical Lie reduction, the number of cases that need to be examined reduces to:
\begin{center}
\begin{tabular}{cp{4cm}p{6cm}c}
\hline
\rule[-3mm]{0mm}{8mm}{\bf 1} & $\langle \DDD, \p_t\rangle$ & $\psi = \sqrt{(x^2+y^2)^3}\,v(\varphi)$ & $\varphi = \arctan{\frac{y}{x}}$ \\
\rule[-3mm]{0mm}{5mm}&\multicolumn{3}{c}{$v(w+\beta\sin\varphi)_\varphi -\frac{1}{3}v_\varphi(w+\beta\sin\varphi)=0, \quad w := v_{\varphi\varphi} + 9v$}\\
\hline
\end{tabular}
\end{center}
The first of the above equations implies the following functional relation between~$w$ and~$v$:
$
    w + \beta \sin\varphi = c_0v^{\frac{1}{3}}.
$
If $c_0=0$, the second equation can be easily integrated with respect to~$v$. This leads to the invariant solution
\[
    \psi = c_1(x^2-3y^2)x + c_2(3x^2-y^2)y-\frac{\beta}{8}(x^2+y^2)y
\]
of \eqref{eq:vortbeta}. 
In the case $c_0\ne 0$, we find particular solutions of the second equation which give rise to the invariant solutions
\[
    \psi = \frac{\beta}{2}(x^2+y^2)^{\frac32}\sin^3\left(\frac{1}{3}\arctan{\frac{y}{x}}\right), \qquad 
    \psi = -\frac{\beta}{2}(x^2+y^2)^{\frac32}\sin^3\left(\frac{1}{3}\arctan{\frac{y}{x}}\pm\frac\pi3\right).
\]

\subsection{Partially invariant solutions}

For a system with at least two dependent variables, it is possible to determine partially invariant solutions \cite{ovsi82Ay}. 
The construction of partially invariant solutions has already been extensively considered in hydrodynamics~\cite{andr98Ay,golo08Ay,hema02Ay,mele04Ay,popo00Ay}. 
In this part, we compute some partially invariant solutions for the \bbve. 
First of all, it is noted that any single equation can be split into a system of multiple equations in various ways 
introducing a new dependent variable for each additional equation desired. 
We consider the \bbve~as the system of two PDEs
\begin{align}\label{eq:vortbeta2}
    \zeta_t + \psi_x\zeta_y - \psi_y\zeta_x + \beta\psi_x = 0, \qquad   \zeta = \psi_{xx} + \psi_{yy},
\end{align}
where both $\psi$ and $\zeta$ are treated as dependent variables.
The splitting of eqn.~\eqref{eq:vortbeta} into system \eqref{eq:vortbeta2} is quite natural 
since both $\psi$ and $\zeta$ have an obvious physical importance. 
Of course, it is not unique. 
Another natural splitting is given by  
the system in terms of the usual velocity variables together with the condition of vanishing divergence. 
However, here we will not pursue any other splittings further.

It is an important property of the chosen splitting that the maximal Lie invariance algebra $\mathcal{B}^\infty_{1\mathrm{s}}$ of \eqref{eq:vortbeta2} is isomorphic 
to the algebra $\mathcal{B}^\infty_1$. More precisely, every operator from $\mathcal{B}^\infty_{1\mathrm{s}}$ is a prolongation of an operator 
from $\mathcal{B}^\infty_1$. This is why for the construction of partially invariant solutions we can use the lists of subalgebras obtained above.

\medskip

As an example for a partially invariant solution, we use the subalgebra $\langle \XX(1), \ZZ(g) \rangle$. Due to the generator $\ZZ(g)$, we cannot make an ansatz for $\psi$. However, we can make an ansatz for $\zeta$ and because of the generator $\p_x$, we have $\psi=\psi(t,x,y)$ and $\zeta = \zeta(t,y)$. Therefore, \eqref{eq:vortbeta2} is reduced to
$
    \zeta_t + \psi_x(\zeta_y + \beta) = 0,$ $\zeta = \psi_{xx} + \psi_{yy}.
$
Introducing the absolute vorticity $\eta = \zeta + \beta y$ and setting $\psi = \Psi(t,x,y) + \tilde \zeta (t,y)$ with $\tilde \zeta_{yy} = \zeta$, we find
$\eta_t + \Psi_x\eta_y = 0,$ and $\Psi_{xx} + \Psi_{yy} = 0$.
If $\eta_y = 0$, we have $\eta_t = 0$ and, consequently, $\eta = \const$. The stream function constructed in this way then reads as
\[
    \psi = \Psi - \frac{1}{6}\beta y^3 + \frac{1}{2}\eta y^2
\]
where $\Psi(t,x,y)$ is an arbitrary solution of the Laplace equation $\Psi_{xx} + \Psi_{yy} = 0$.

In case $\eta_y \ne 0$, we find that the stream function has the form
\[
    \psi = \frac{1}{(g^1)^2}F(\omega) - \frac{1}{6}\beta y^3 - \frac{g^1_ty+g_t^0}{g^1}x + f^1y + f^0,
\]
where $\omega = g^1y + g^0$ and $g^1,g^0,f^1$, and $f^0$ are functions of $t$.

\medskip

To present one more example of a partially invariant solution, we take the subalgebra $\langle \p_y, \ZZ(g)\rangle$. Similar to the previous case, we now have $\zeta = \zeta(t,x)$ and $\psi = \psi(t,x,y)$. Then, \eqref{eq:vortbeta2} is reduced to
\mbox{$
    \zeta_t -\psi_y\zeta_x + \beta\psi_x = 0$} and  $\zeta = \psi_{xx}+\psi_{yy}.
$
Introducing $\psi = \phi(t,x,y) + \sigma(t,x)$, $\sigma_{xx} = \zeta$, this set of equations yields
\begin{align} \label{eq:vortbetapis}
    \phi_{xx} + \phi_{yy} = 0, \qquad   \sigma_{xxt} - \sigma_{xxx}\phi_y + \beta(\phi_x+\sigma_x) = 0.
\end{align}
We now have to distinguish different cases for the integration of this system.

\medskip

\noindent {\bf 1.} $\sigma_{xxx} = 0$. In this instance, the solution for the stream function reads as
\[
    \psi = -\frac{1}{\beta}(2\chi^1_tx+\chi^2)+ \chi^1y^2+\chi^3y,
\]
where $\chi^1$, $\chi^2$ and $\chi^3$ are smooth functions of~$t$. 
The functions $\chi^2$ and $\chi^3$ can be set equal to 0 by symmetry transformations generated by an operator of the form $\XX(f)+\ZZ(g)$.

\medskip

\noindent {\bf 2.} $\sigma_{xxx} \ne 0$. We set
$
    \phi = H-\beta^{-1}\sigma_{xt} - \sigma
$
and substitute into the second equation of \eqref{eq:vortbetapis}, which yields a characteristic system for $H$. Solving this system, we find that $H=H(t,\eta)$, where $\eta=\sigma_{xx}+\beta y$ is again the absolute vorticity. From the first equation of \eqref{eq:vortbetapis}, we then derive
\begin{equation}\label{eq:pis0}
    H_{\eta\eta} ((\sigma_{xxx})^2+\beta^2) + H_\eta\sigma_{xxxy} - \frac{\sigma_{xxxt}}{\beta} - \sigma_{xx}  = 0.
\end{equation}
If we fix $x$ in the above equation, we can write
$
    h^2(t)H_{\eta\eta} + h^1(t)H_\eta + h^0(t) = 0.
$
We now have to distinguish whether there are two independent equations of this type or only one.

In case of two equations we have $H_{\eta\eta}=0$ and consequently $H=\alpha(t)\eta+\gamma(t)$. Substituting this into \eqref{eq:pis0}, solving the resulting PDE and transforming back to the original variables, we find
$
    \psi = \Sigma(t,x) + \alpha(t)\beta y,$ $\alpha\Sigma_{xx} - \beta^{-1}\Sigma_{xt} + \beta^{-1}\delta(t) - \Sigma = 0.
$
By means of symmetry transformations generated by $\XX(f)$ and $\ZZ(g)$ we can set $\alpha=\delta=0$ and again arrive at the Klein-Gordon equation. This illustrates the fact that in some cases the ansatz for a partially invariant solution effectively reduces to a usual group-invariant reduction.

If we only have one independent equation in~$H$, $h^2\ne0$, and the equation
\[
    H_{\eta\eta} ((\sigma_{xxx})^2+\beta^2) + H_\eta\sigma_{xxxx} - \frac{\sigma_{xxxt}}{\beta} - \sigma_{xx}  = \lambda(h^2(t)H_{\eta\eta} + h^1(t)H_\eta + h^0(t)),
\]
where $\lambda=\lambda(t,x)$, holds identically in~$H$. Splitting this equation with respect to~$H$ leads to the three equations
\[
(\sigma_{xxx})^2+\beta^2=\lambda h^2,\qquad  \sigma_{xxxx}=\lambda h^1,\qquad -\beta^{-1}\sigma_{xxxt}-\sigma_{xx} = \lambda h^0.
\]
Since $h^2\ne0$, we can express $\lambda$ from the first equation. 
Provided that $h^1=0$, we integrate the second equation to find
$
    \sigma = \sigma^3(t)x^3+\sigma^2(t)x^2+\sigma^1(t)x+\sigma^0(t).
$
Inserting this expression in the third equation and splitting with respect to $x$ then yields
$\sigma^3=0$, i.e., $\sigma_{xxx} = 0$, contradicting the initial assumption for this case.
For $h^1\ne0$, we integrate the second equation once with respect to $x$ and then substitute the resulting expression for $\sigma_{xx}$ into the third equation. This leads to a contradiction in the system constructed by splitting with respect to~$x$, and hence no solution is obtained also under the assumption $h^1\ne0$.

\section{The spherical equation}

\subsection{The model}

The barotropic vorticity equation on the sphere (\sbve) is given by (e.g.\ Ref. \cite{plat60Ay})
\begin{equation}\label{eq:vortsphere}
\zeta_t + \frac{1}{R^2}\left(\psi_\lambda\zeta_\mu - \psi_\mu\zeta_\lambda\right) + \frac{2\Omega}{R^2}\psi_\lambda = 0,\quad 
\zeta := \frac{1}{R^2}\left[\frac{1}{1-\mu^2}\psi_{\lambda\lambda} + \left((1-\mu^2)\psi_\mu\right)_\mu\right],
\end{equation}
where $\psi$ is the (spherical) stream function and $\zeta$ the (spherical) vorticity.
\looseness=-1
They are related through the Laplacian on the sphere. Instead of using the latitude $\varphi$ as an independent variable, in practice, it is convenient to rather use $\mu = \sin\varphi$. The value of $\mu$ ranges from $-1$ (South Pole) to $1$ (North Pole). By~$\lambda$ we denote the longitude, $R$ is the mean radius of the Earth and $\Omega$ the absolute value of the Earth's angular rotation vector.

\subsection{The symmetries}\label{sec:SymmetriesSphericalVersion}

We aim to start with~\eqref{eq:vortsphere} in a nonrotating reference frame ($\Omega = 0$). Note in passing that it is possible to scale the radius~$R$ of the Earth to~1 by including~$R$ in the stream function via setting $\tilde \psi = \psi / R^2$.

The corresponding Lie symmetry algebra $\mathcal{S}_0^\infty$ is infinite dimensional and a suitable basis is provided by
\begin{align}\label{eq:gensphere}
\begin{split}
    &\mathcal{D} = t\p_t - \psi\p_{\psi},\qquad \p_t,\qquad \mathcal{Z}(g) = g(t)\p_{\psi},\qquad \JJ_1 = \p_{\lambda}, \\
    &\JJ_{2} = \mu\frac{\sin \lambda}{\sqrt{1-\mu^2}} \p_{\lambda} + \sqrt{1-\mu^2}\cos\lambda\p_{\mu},\qquad\JJ_{3} = \mu\frac{\cos \lambda}{\sqrt{1-\mu^2}} \p_{\lambda} - \sqrt{1-\mu^2}\sin\lambda \p_{\mu}.
\end{split}
\end{align}
As for the physical meaning of these basis elements, we find that $\mathcal{D}$ is the generator of scaling in $t$ and $\psi$ and $\p_t$ corresponds to time translations. The generators $\mathcal{J}_{i}$, $i=1,2,3$, correspond to rotations in angular coordinates. This follows since they satisfy the commutation relations of the Lie algebra $\mathfrak{so}(3)$, $[\JJ_i,\JJ_j] = \sum_{k=1}^3\varepsilon_{ijk}\JJ_k$, where $i,j=1,2,3$ and $\varepsilon_{ijk}$ is the Levi--Civita symbol. $\mathcal{Z}(g)$ again represents gauging of the stream function.

The algebra $\mathcal{S}_0^\infty$ has the structure of $\mathfrak{so}(3)\oplus(\mathfrak{g}_2\lsemioplus \langle\ZZ(g)\rangle)$, where $\mathfrak{g}_2 = \langle \DDD, \p_t\rangle$ is the two-dimensional nonabelian algebra and $\langle\ZZ(g)\rangle$ is an infinite-dimensional Abelian ideal in $\mathcal{S}_0^\infty$.

Now turning to the rotating case ($\Omega \ne 0$). Eqn.\ \eqref{eq:vortsphere} admits the infinite-dimensional Lie invariance algebra~$\mathcal{S}_\Omega^\infty$
\begin{align*}
     &\DDD = t\p_t - (\psi-\Omega\mu)\p_{\psi} - \Omega t\p_{\lambda}, \qquad \p_t, \qquad \ZZ(g) = g(t)\p_{\psi},\qquad \JJ_{1} = \p_{\lambda}, \\
     &\JJ_{2} = \mu\frac{\sin(\lambda+\Omega t)}{\sqrt{1-\mu^2}}\p_{\lambda}+\frac{\cos(\lambda+\Omega t)}{\sqrt{1-\mu^2}} \left((1-\mu^2)\p_{\mu} + \Omega\p_{\psi}\right), \\ 
     &\JJ_{3} =\mu \frac{\cos(\lambda+\Omega t)}{\sqrt{1-\mu^2}}\p_{\lambda}-\frac{\sin(\lambda+\Omega t)}{\sqrt{1-\mu^2}} \left((1-\mu^2)\p_{\mu} + \Omega\p_{\psi}\right).    
\end{align*}
The physical interpretation of the basis elements is obvious from those of the case $\Omega=0$. Moreover, straightforward calculation shows that both Lie symmetry algebras $\mathcal{S}_0^\infty$ and $\mathcal{S}_\Omega^\infty$ are isomorphic and can be mapped to each other by means of the change in the coordinates,
\begin{equation}\label{eq:vorttrans}
    \tilde t = t, \quad \tilde\mu = \mu, \quad \tilde\lambda = \lambda + \Omega t, \quad \tilde\psi = \psi - \Omega\mu.
\end{equation}
Furthermore, it is possible to transform \eqref{eq:vortsphere} into the corresponding equation in the rest frame $(\Omega = 0)$ upon using \eqref{eq:vorttrans}. This recovers, in a systematic way, the transformation used by Platzman \cite{plat60Ay} to reduce the spherical vorticity equation to a reference frame with zero angular momentum.

Note that this mapping is possible due to the special form of the Laplacian in spherical coordinates. In particular, it is impossible to obtain a similar result for the vorticity equation in Cartesian coordinates since in this case the respective Lie symmetry algebras are nonisomorphic. Consequently, no transformation can be found that maps the vorticity equation on the $\beta$-plane to the vorticity equation on the $f$-plane. This indicates that the traditional $\beta$-plane approximation significantly distorts the geometry of the more natural spherical vorticity dynamics.

Again there are two discrete symmetries, given by $(t,\lambda,\mu,\psi)\mapsto(-t,-\lambda,\mu,\psi)$ and $(t,\lambda,\mu,\psi)\mapsto(t,\lambda,-\mu,-\psi)$, respectively.

\subsection{Classification of subalgebras}

The classification of subalgebras of $\mathcal{S}_0^\infty$ is done in the same fashion as for the $\mathcal{B}^\infty_1$. The nonidentical adjoint actions involving basis elements of the algebra $\mathcal{S}_0^\infty$ are exhausted by the following list:
\begin{align*}
    &\Ad(e^{\ve\p_t})\DDD = \DDD - \ve\p_t  & &\Ad(e^{\ve\DDD})\p_t = e^\ve\p_t \\
    &\Ad(e^{\ve\ZZ(g)})\DDD = \DDD + \ve\ZZ(g+tg') &&\Ad(e^{\ve\ZZ(g)})\p_t = \p_t + \ve\ZZ(g') \\
    &\Ad(e^{\ve\JJ_{1}})\JJ_2 = \phantom{-}\JJ_2\cos\ve + \JJ_3\sin\ve  &&\Ad(e^{\ve\JJ_{2}})\JJ_3 = \phantom{-}\JJ_3\cos\ve + \JJ_1\sin\ve \\
    &\Ad(e^{\ve\JJ_{1}})\JJ_3 = -\JJ_2\sin\ve + \JJ_3\cos\ve && \Ad(e^{\ve\JJ_{2}})\JJ_1 = -\JJ_3\sin\ve + \JJ_1\cos\ve \\
    &\Ad(e^{\ve\JJ_{3}})\JJ_1 = \phantom{-}\JJ_1\cos\ve + \JJ_2\sin\ve & &\Ad(e^{\ve\p_t})\ZZ(g) = \ZZ(\tilde g), \quad \tilde g = g(t-\ve)\\
    &\Ad(e^{\ve\JJ_{3}})\JJ_2 = -\JJ_1\sin\ve\ + \JJ_2\cos\ve  &&\Ad(e^{\ve\DDD})\ZZ(g) = \ZZ(\tilde g), \quad \tilde g = e^{-\ve}g(e^{-\ve}t).
\end{align*}

Similar to the case of the \bbve, we start with the most general form of an infinitesimal generator
$
    \vv = a_D\DDD + a_t\p_t + a_1\JJ_1 + a_2\JJ_2 + a_3\JJ_3 + \ZZ(g).
$
In the same manner, by acting with the adjoint actions given above, we can determine the following list of conjugacy inequivalent \emph{one-dimensional subalgebras} of $\mathcal{S}_0^\infty$:
\begin{align}\label{eq:classificationsphere}
    \langle \DDD +a\JJ_1\rangle,\qquad \langle \p_t+a\JJ_1\rangle, \qquad \langle\JJ_1 + \ZZ(g) \rangle,\qquad \langle \ZZ(g) \rangle,
\end{align}
where $a\in \mathbb{R}$ and $a\in \{-1,0,1\}$ for the first and second cases, respectively. Unlike the case of the \bbve, there is no discrete symmetry allowing placement of additional restrictions on the values of $a$.
There are equivalence relations within the two last families of subalgebras,  
generated by adjoint actions of the scaling transformations, time translations, and within the last family, changes of algebra bases. 

Using the same procedure as described in the second part of section \ref{sec:class2beta} we find the following list of conjugacy inequivalent \textit{two-dimensional subalgebras} of \eqref{eq:gensphere}:
\begin{gather*}
    \langle \DDD + a\JJ_1, \p_t \rangle, \qquad \langle \DDD, \JJ_1 + \ZZ(at^{-1})\rangle, \qquad \langle \DDD + a\JJ_1, \ZZ(|t|^b)\rangle, \\ 
    \langle\p_t, \JJ_1 + \ZZ(c) \rangle, \qquad \langle\p_t + c\JJ_1, \ZZ(e^{\tilde ct}) \rangle, \qquad 
    \langle\JJ_1 + \ZZ(g^1), \ZZ(g^2) \rangle, \qquad  \langle\ZZ(g^1), \ZZ(g^2) \rangle,
\end{gather*} 
where $a,b\in\mathbb R$, $c\in \{-1,0,1\}$; $\tilde c\in \{-1,0,1\}$ if $c=0$. 
There are additional equivalence relations within the last two series of subalgebras,  
generated by adjoint actions of the scale transformations, time translations and changes in the algebra bases.

\subsection{Group-invariant reduction with one-dimensional subalgebras}

Based on the above classification of one-dimensional algebras, below we present the corresponding list of reduced differential equations obtained from the \sbve\ here. Again, $p,q$ denote the new independent variables, while $v=v(p,q)$ is the new dependent variable.

\begin{center}
\begin{tabular}{cp{4cm}p{6cm}c}
\hline
\rule[-3mm]{0mm}{8mm}{\bf 1} & $\langle\DDD + a\JJ_1 \rangle$ & $\psi = t^{-1}v$ & $p = \lambda -a\ln t,\quad q = \mu$ \\
\rule[-3mm]{0mm}{5mm}&\multicolumn{3}{c}{$w + aw_p - v_pw_q + v_qw_p = 0, \quad w := \tfrac{1}{1-q^2}v_{pp} + ((1-q^2)v_q)_q$}\\
\hline
\rule[-3mm]{0mm}{8mm}{\bf 2} & $\langle \p_t+a\JJ_1\rangle$ & $\psi = v$ & $p = \lambda - at,\quad q = \mu$ \\
\rule[-3mm]{0mm}{5mm}&\multicolumn{3}{c}{$-(aq+v)_qw_p + (aq+v)_pw_q = 0, \quad w := \frac{1}{1-q^2}v_{pp} + ((1-q^2)v_q)_q$} \\
\hline
\rule[-3mm]{0mm}{8mm}{\bf 3} & $\langle\JJ_1 + \ZZ(g) \rangle$ & $\psi = v + g(t)\lambda$ & $p = t,\quad q = \mu$ \\
\rule[-3mm]{0mm}{5mm}&\multicolumn{3}{c}{$w_p + gw_q = 0, \quad w := ((1-q^2)v_q)_q$}\\
\hline
\rule[-3mm]{0mm}{8mm}{\bf 4} & $\langle \ZZ(g) \rangle$ & \multicolumn{2}{c}{No group-invariant reduction is possible in this case} \\
\hline
\end{tabular}
\end{center}

All Lie symmetries of the reduced equations of Cases~{\bf 1} and~{\bf 2} are induced by Lie symmetries of the~\sbve.
This is why further Lie reductions of these cases give no new solutions in comparison to Lie reductions with respect to two-dimensional
subalgebras.

We now give some examples for solutions obtained upon using the above ans\"atze:

\medskip

\noindent {\bf Case 2} includes the well-known Rossby--Haurwitz wave solutions. To show this, we construct a class of exact solutions upon using invariance of the \sbve\ under the algebra $\langle \p_t+a\JJ_{1}\rangle$. In particular, the corresponding reduced vorticity equation implies that $w = F$, where $F$ is a function of $v+aq$. Hence, we have
\begin{equation}\label{eq:vorthaurwitz}
    \frac{1}{1-q^2}v_{pp} + (1-q^2)v_{qq} - 2q v_q = F.
\end{equation}
Eqn.\ \eqref{eq:vorthaurwitz} is, in general, a nonlinear Poisson equation in spherical coordinates. To obtain the Rossby--Haurwitz wave solution from this equation, we set $F = c(v+aq)$, $c=\const$, that is, we make a homogeneous linear ansatz for $F$. Separation of the variables gives the ansatz
\[
        v(p,q) = Ae^{imp}P_n^m(q) + Be^{imp}Q_n^m(q) - \frac{acq}{c+2},
\]
with $A,B=\const$, where $P^m_n(q)$ and $Q^m_n(q)$ are the associated Legendre functions of the first and second kind, respectively, and the degree $n$ is given by
\begin{equation}\label{eq:cond}
    n = \frac{1}{2}(\sqrt{1-4c}-1).
\end{equation}
For the sake of brevity, we now set $B=0$. Reverting to the original variables and employing transformation \eqref{eq:vorttrans} to map the solution of the \sbve\ with vanishing rotation to a solution of the \sbve\ with rotation, we find
\begin{equation}\label{eq:vorthaurwitzsol}
    \psi(t,\lambda,\mu) = AP_n^m(\mu)e^{im(\lambda-(a-\Omega)t)} - \frac{ac\mu}{c+2}+\Omega\mu.
\end{equation}
To derive pure wave solutions, we require $a = \Omega(c+2)/c$, which, upon inserting in \eqref{eq:vorthaurwitzsol} and considering \eqref{eq:cond}, allows us to arrive at the well-known phase relation for a single Rossby--Haurwitz wave (e.g. Refs. \cite{haur40Ay,neam46Ay,plat60Ay}):
\begin{equation}\label{eq:haurwitzphase}
    c_{\textup{phase}} := a - \Omega = -\frac{2\Omega}{n(n+1)}.
\end{equation}
Since the integer constant $m$ is arbitrary and only linear cases of eqn.~\eqref{eq:vorthaurwitz} are considered, we may extend the solution \eqref{eq:vorthaurwitzsol} by superposition of single solutions with $m$ ranging from $-n$ to $n$. This recovers---upon using \eqref{eq:haurwitzphase}---the classical ansatz for the stream function of Rossby--Haurwitz waves.

Moreover, note that the class of solutions that may be obtained upon employing symmetry methods is again much wider than that obtained upon using the usual ansatz for the stream function. In fact, it can be seen that Rossby--Haurwitz waves correspond to particular simple solutions of the reduced spherical vorticity equation \eqref{eq:vorthaurwitz}, but there is an infinite class of other solutions invariant under the same generator $\p_t + a\JJ_{1}$.

\medskip

\noindent {\bf Case 3} is completely integrable by quadratures. The general solution for the stream function reads
\[
    \psi = g(t)\lambda + f(t) + h(t)\textup{arctanh} \mu + \int \frac{\int w(\theta) \ddd \mu}{1-\mu^2}\ddd \mu, \qquad
    \theta:=\mu-\int g(t){\rm d}t.
\]

\subsection{Group-invariant reduction with two-dimensional subalgebras}

As was discussed in section \ref{sec:red2beta} it is, in general, not necessary to investigate reductions with the complete set of two-dimensional inequivalent subalgebras. 
Namely, if some of the equations obtained from one-dimensional reduction are completely integrable, 
we can avoid the computation of reduction with two-dimensional subalgebras if these algebras contain the generators that enabled the 
complete integration in the first place. 
For the \sbve, case {\bf 3} is integrable and case {\bf 4} does not allow to compute classical group-invariant solutions. 
Hence, we again have only the reduction in one two-dimensional subalgebra, which is not trivial in view of the reductions based on one-dimensional subalgebras: $\langle \DDD + a\JJ_1, \p_t\rangle$. 
Note, however, that it is not possible to use this subalgebra in the case $a=0$ 
for a classical Lie reduction since no proper ansatz for $\psi$ can be constructed. 
Rather, it can only be used for the construction of partially invariant solutions. 
If $a\ne0$, this subalgebra leads to invariant solutions that are obtainable as particular cases of reduction with the algebra $\langle \p_t\rangle$. 
The corresponding ansatz $\psi=e^{b\lambda}v(\mu)$ 
reduces the \sbve\ to the equation $vw_\mu-v_\mu w=0$, where $b=-1/a$, $w:=b^2(1-\mu^2)^{-1}v + ((1-\mu^2)v_\mu)_\mu$.
The reduced equation implies the following linear constraint between $v$ and~$w$: $w=Cv$, where $C$ is an arbitrary constant, i.e., we have the equation
\[
((1-\mu^2)v_\mu)_\mu+\frac{b^2}{1-\mu^2}v=Cv
\]
which is integrable in terms of Legendre functions.

\subsection*{Acknowledgements}

This research was supported by the Austrian Science Fund (FWF), project No.\ P20632. A.B.\ is a recipient of a DOC-fellowship of the Austrian Academy of Sciences.
The authors are grateful to Professor M.~Kunzinger and Professor A.~Sergyeyev for productive and helpful discussions and thank the referee for useful remarks.

\end{document}